# Twistor diagram recursion for all gauge-theoretic tree amplitudes

Andrew Hodges

*Wadham College, University of Oxford, Oxford OX1 3PN, United Kingdom*

March 2005

*Abstract:* The twistor diagram formalism for scattering amplitudes is introduced, emphasising its finiteness and conformal symmetry. It is shown how MHV amplitudes are simply represented by twistor diagrams. Then the Britto-Cachazo-Feng recursion formula is translated into a simple rule for composing twistor diagrams. It follows that all tree amplitudes in pure gauge-theoretic scattering are expressed naturally as twistor diagrams. Further implications are briefly discussed.

**1. Introduction**

Recent work has greatly advanced the study of quantum-field-theoretic amplitudes with elegant and powerful techniques focused on N=4 supersymmetric gauge field theory. A frequent comment is that the results turn out to be much simpler than expected, and that there must be some more fundamental structure yet to be elucidated. These recent advances have been stimulated by the new formulation of twistor-string theory by Witten (2003), which indicated *twistor space* as the correct setting for this more fundamental structure. Yet twistor geometry has not played a central role in the exposition of these very recent advances. In particular, the elegant and powerful *recursion relation* of Britto, Cachazo and Feng (2004) for building gauge-theoretic amplitudes is formulated and proved without any explicit reference to twistor geometry. Therefore it has been doubted whether twistor theory actually plays an essential role in the advances being made in gauge-field theory.

The purpose of this note is to point out that in fact, the BCF recursion is intimately related to twistor theory. It can be naturally and simply stated in terms of the representation of amplitudes by *twistor diagrams*. Their recursion formula is equivalent to a simple graph-theoretic rule for joining twistor diagrams together. Thus the new recursion relation, far from indicating the possible irrelevance of twistor geometry, suggests that it may be crucial to further progress.

Twistor diagrams were originally defined by Roger Penrose in about 1970 as the analogue in twistor space of Feynman diagrams in space-time. The formalism proposed by Penrose has since then been radically modified, but it retains the most vital original characteristics. As with Feynman diagrams, twistor diagrams are integrals built out of simple standard components. They are entirely holomorphic, using only contour integration. They make explicit where *conformal symmetry* holds and where it is broken. They are gauge-invariant. Moreover they are *manifestly finite,* the contours all being compact. This finiteness is achieved through a regularising principle which is essentially twistor-geometric.

As we shall show, Britto, Cachazo and Feng's formula ensures that all tree amplitudes are simply expressible as twistor diagrams. However, there is no reason to suppose that a restriction to tree amplitudes is essential, and we shall add some remarks on the prospects for representing loop amplitudes. We also point out some wider implications of this reformulation.

## 2. Twistor Diagrams

As twistor diagrams are not well known, and as the last review (Hodges 1998) is rather out of date, the formalism will be very briefly introduced. This is best done by example, rather than by abstract definition. The natural starting-point is supplied by exhibiting the twistor diagrams for the scattering of *four* gauge fields. We shall assume throughout a general knowledge of how the gauge-field amplitudes separate into a sum of 'colour-stripped' amplitudes. We first state the twistor diagram for the colour-stripped amplitude $A(1^- \ 2^+ \ 3^- \ 4^+)$:

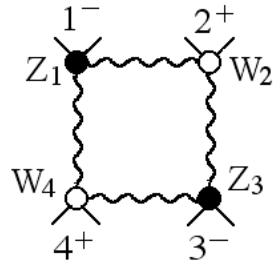

The vertices are the analogues of Feynman diagram vertices, but instead of each representing an integral over momentum space, each represents an integral in twistor space. More precisely, each black vertex ($Z_1$, $Z_3$ in the example) represents an integral over a twistor variable, and each white vertex an integral over a *dual* twistor variable ($W_2$, $W_4$). The twistor and dual twistor spaces are simply copies of $\mathbf{C}^4$, and the differential forms simply composed of the various $d^4Z$, $d^4W$, corresponding to each vertex. The contours are 16-real-dimensional, inside the 16-complex-dimensional product space. The outward-pointing lines indicate external *gauge fields* attached to the vertices, in their twistor or dual twistor representation. In this *particular* diagram these are homogeneous of degree ($-4$) at each vertex. Thus, they represent the ($-+-+$) ordering of helicities. However, the external fields should *not* be thought of as simply forming a product of the four free fields, each represented by a first-cohomology element on twistor or dual twistor space. They should be thought of as a four-field state in the Fock space, requiring some extra structure, and so composing (in this example) a four-twistor function $F(Z_1, W_2, Z_3, W_4)$. We shall briefly mention something about this extra structure shortly.

It is important that each space is a **C**$^4$ and *not* a **CP**$^3$. This is because the wavy lines joining the vertices, which may be called *line-propagators,* have the following definition: each one constrains the contour to have a *boundary* on a subspace of form

$$W_\alpha Z^\alpha = k$$

which is essentially *inhomogeneous*. Putting all this together, what this diagram actually means is just the very simple integral:

$$\int_{\substack{W_2 \cdot Z_1 = k, W_2 \cdot Z_3 = k \\ W_4 \cdot Z_1 = k, W_4 \cdot Z_3 = k}} F(Z_1, W_2, Z_3, W_4) \, d^4Z_1 d^4W_2 d^4Z_3 d^4W_4$$

but such an expression is not very helpful because all the important content resides in the location of the boundaries of the contour, i.e. the geometric shape of the region of integration, and it is not a good idea to cram this content into the subscript of the integral sign. This is just one of many reasons why the diagram notation is useful. All questions of numerical factors and overall sign are being ignored here, although precision will be necessary, along with a full definition of oriented contours, in a complete theory.

What has this integral got to do with the Parke-Taylor formula which expresses the correct amplitude? Actually, it can be thought of as a *conformally invariant version* of that formula, which would normally be written as

$$A(1^- 2^+ 3^- 4^+) = \frac{\langle 13 \rangle^4}{\langle 12 \rangle \langle 23 \rangle \langle 34 \rangle \langle 41 \rangle} \delta(p_1 + p_2 + p_3 + p_4)$$

To see why this is, we need first to note that the spinor product $\langle 12 \rangle$ corresponds to the operator $I_{\alpha\beta} Z_1^\alpha \, \partial / \partial W_2^\beta$, where $I_{\alpha\beta}$ is the antisymmetric *infinity twistor* which picks out the momentum-spinor parts out of twistors. This is a simple consequence of the twistor quantization which gives rise to the representation of fields by homogeneous twistor functions. Applying this operator, the fundamental theorem of calculus implies that the boundary becomes a *simple pole* and we obtain

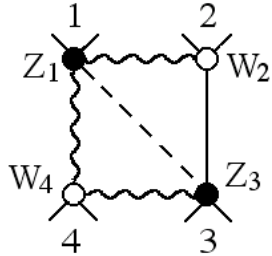

where the new single line between $W_2$ and $Z_3$ represents the simple pole

$$\left(W_{2\alpha} Z_3^\alpha - k\right)^{-1}.$$

The dashed line represents the *numerator factor* $I_{\alpha\beta} Z_1^\alpha Z_3^\beta$. Although at first laborious, it will be found that the diagram formalism makes all these operations simple to apply. Continuing in this way we obtain an integral which is indeed of the form of $\langle 13 \rangle^4$ multiplying a new integral. For agreement with the Parke-Taylor formula, this new integral must be the δ-function on four scalar fields. It is:

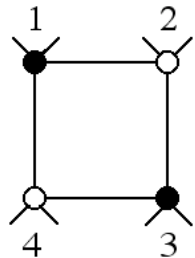

But this twistor integral indeed does correspond to the requisite δ-function: the correspondence has been studied in twistor theory from early days (Penrose 1972, 1975). To see why, here is an intuitive argument, which can be made exact. The inhomogeneous $k$ actually makes no difference to the answer in this case (expand as a Taylor series in positive powers of $k$: by homogeneity, only the leading term is non-vanishing.) Hence in this case the result can be imitated by a projective integral. The four simple poles then have the effect of restricting all the variables on to a common *line* in twistor space. Making a correct choice of contour is equivalent to letting this line vary over the points of (compactified) Minkowski space, and so effects the same as integrating the four scalar fields, evaluated and multiplied together at a common point, over space-time. This is exactly what a δ-function in momentum space means.

We have not introduced momentum states. Instead, we have have followed the lead taken in (Penrose 1972), and have derived everything by first representing the δ-function, and then applying helicity-changing differential operators. In so doing we have made use of the principles of twistor quantization in which everything becomes holomorphic. From this point of view, the question of 'real' momenta does not really arise inside the diagram: momenta are holomorphic differential operators. The formalism automatically absorbs the relations between the momenta implied by the δ-function.

The diagram we have considered is necessarily asymmetric, in that a twistor representation is used for two external fields, and a dual twistor for the other two. But if we wished we could make all the external fields into functions on *twistors,* by attaching *twistor transforms* to the $W_2$, $W_4$ vertices and considering the diagram

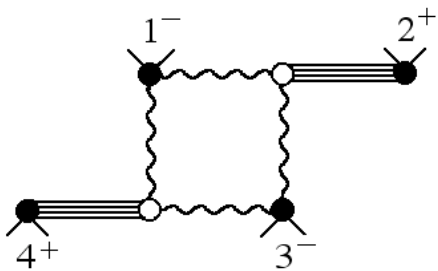

where the new lines represent *quadruple poles* of form $-6\left(W_\alpha Z^\alpha - k\right)^{-4}$.

Now the corresponding external twistor functions must be homogeneous of degree 0 in $Z_2$ and $Z_4$. The $\langle 12 \rangle$ operator now corresponds to $I_{\alpha\beta} Z_1^\alpha Z_2^\beta$. Application of integration by parts yields the same results as before.

In future development of the theory it may become crucial to impose that the external fields are (say) functions of twistors rather than dual twistors. But for present purposes it is not necessary to make any choice and the lines pointing to external fields will be deliberately left vague in the diagrams drawn in this note.

There is in any case another reason for this vagueness, which is that the specification given above is not yet the whole story. Although formally we have a representation of the Parke-Taylor amplitude, as shown above, the elements as so far given are insufficient to ensure the actual existence of *contours* for all channels. Another element is needed for this: the integral must be allowed further boundaries on subspaces of form $I_{\alpha\beta} X^\alpha Z^\beta = M$ or $I^{\alpha\beta} W_\alpha Y_\beta = M$, where $M$ has the dimensions of a mass. Of course, these boundaries *break the conformal invariance*. It has only recently been realised that these boundaries are essential not just for obtaining scattering amplitudes, but for the very concept of an entangled many-field quantum state. As mentioned earlier, the external $F(Z,W.....)$ should not be thought of simply as a product of functions (more precisely, first-cohomology elements) on twistor spaces, but as elements in a larger product space. It appears that these elements must in fact be defined through *relative cohomology* which brings in just these inhomogeneous and conformal-symmetry breaking boundaries. The operators of form $I_{\alpha\beta} \, \partial / \partial W_2^\beta$ etc do not 'see' these boundaries at all, and all the results to be given this note are unaffected by these considerations governing the representation of the external fields. (A more extensive analysis of the external elements and the contours, etc., will appear later.)

The breaking of conformal invariance is of a minimal nature: it is confined to the effects of these boundaries (which correspond roughly to specifying which are the finite points of space-time). Likewise, the only role of non-holomorphic structure is in specifying the domain of holomorphy for the holomorphic functions defining the external fields, which specifies which are the real points of space-time, and distinguishes past from future.

If evaluation is attempted on *finite-normed elements* of the Hilbert space (which momentum states are *not!*), the collinear singularities of the Parke-Taylor formula yield a divergent integral. For comparison with experiment the representation in terms of momentum states is generally the desired one, so there are good reasons for ignoring this problem for practical purposes. However, the twistor diagram programme has always set out to compute *completely finite* amplitudes, consistent with the fundamentals of quantum mechanics. The inhomogeneous $k$ and $M$ have the remarkable effect of achieving this finiteness: that is, contours exist for the inhomogeneous integral given above, which have no analogue in the corresponding projective integral obtained by letting $k$ and $M$ vanish. It must be stressed that *no* change has thereby been made in the Parke-Taylor formula: it is rather that a more *exact and finite* specification of its meaning has been made through a new sort of boundary condition.

The $k$ and $M$ obviously cannot be represented in *projective* twistor space. This is how Penrose's original proposal, which was for projective twistor integrals, has been most drastically modified (Hodges 1985). Since space-time corresponds to projective twistor space, we are using a regularisation that is not expressible in space-time: it uses an extra (but natural) twistor dimension. Possibly, the effect of this regularisation in loop diagrams will be the same as is obtained by conventional dimensional regularisation. But this cannot be assumed. The twistor regularisation is essentially different in nature, being completely finite and not requiring any limiting operations.

There is of course a tantalising possibility here, which deserves exploration, that the *inhomogeneity* is intimately related to *supersymmetry*. It also seems possible that by generalising this construction, the *gauge group* could play a direct role in the theory. However, for the moment we are writing down the simplest possible version of this kind of inhomogeneous twistor propagator, with a simple scalar and classical number $k$.

There is a natural value for $k$, namely $\exp(-\gamma)$, where $\gamma$ is Euler's constant. There is no obvious value for $M$, nor is there any reason for it to be either small or large.

When amplitudes are evaluated for finite-normed fields, the collinear or infra-red singularity shows up in term involving log($M/k$). This leads to an important aspect of the diagram. By *removing* a boundary line one obtains a *period* or *cut* amplitude. Essentially, it manifests itself as the period of that log($M/k$), and corresponds to an interaction of one order lower. In the case of the process discussed above, there are actually *two* periods, corresponding to the two different zeroth order (no-interaction) processes represented respectively by:

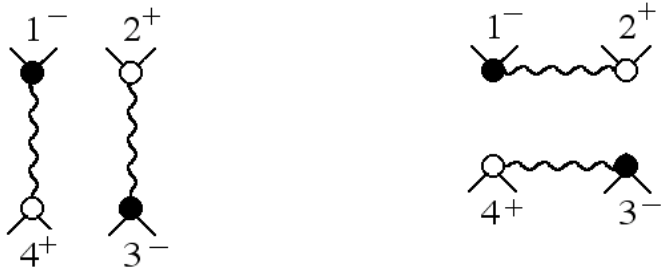

In contrast, the other four-field amplitude A(1⁻ 2⁻ 3⁺ 4⁺) can be represented by either of the twistor diagrams:

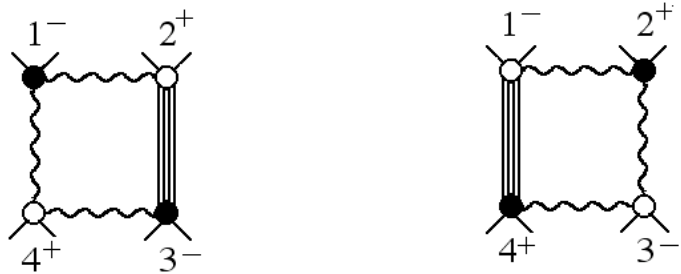

It can be seen immediately from the diagram representation that it has only *one* 'cut' process. The correspondence of these diagrams to the Parke-Taylor amplitude formula

$$A(1^- 2^- 3^+ 4^+) = \frac{\langle 12 \rangle^3}{\langle 23 \rangle \langle 34 \rangle \langle 41 \rangle} \delta(p_1 + p_2 + p_3 + p_4)$$

can be shown by exactly the same methods as used above. The non-uniqueness of these diagram representations is a first taste of a general feature which we shall address later.

These gauge-field diagrams thus combine line-propagators which are either *boundaries* or *quadruple poles*. As we shall see, this simple feature persists for all the tree amplitudes. These line-propagators can be interpreted as helicity transfers of ±1, roughly analogous to the momentum transfers in Feynman diagrams. They can also be regarded as summations over on-shell spin-1 states, *and* as something like square roots of Feynman propagators. The boundary lines are effecting an essential role in *integrating:* both in the Feynman propagator, and in taking the external field information and converting it into a potential. It seems astonishing that all this can be done by just such simple elements. The inhomogeneity means that there is a subtle difference from any definition made in space-time. The full meaning of this difference is not yet clear.

## 3. Delta-function formulas for higher order diagrams

We now wish to obtain diagrams for higher-order processes, proceeding by analogy with the discussion of four-field processes. We must first have suitable representations of the delta-function for any number of scalar fields. Then we can apply helicity-changing operations to obtain corresponding results for gauge-fields.

For the δ-function on *five* scalar fields, we can start with the four-field structure obtained by adding a twistor transform to one vertex:

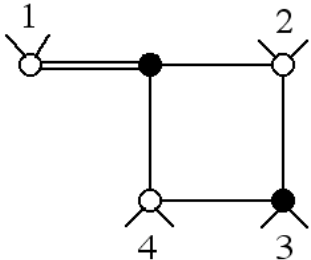

We can then generalise it by multiplying it by the fifth field, evaluated on the line defined by $W_{1\alpha} W_{4\beta}$.

Hence we obtain a five-scalar-field δ-function diagram:

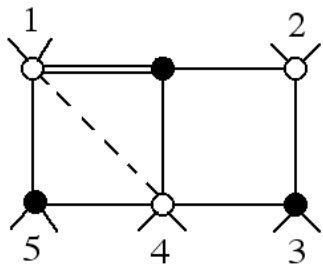

Similarly, for six scalar fields we have:

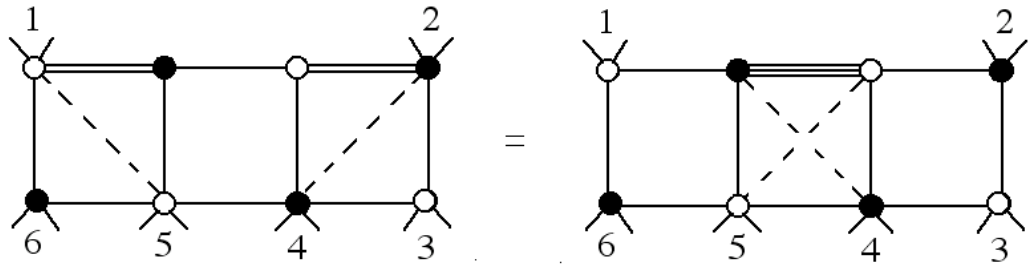

but also

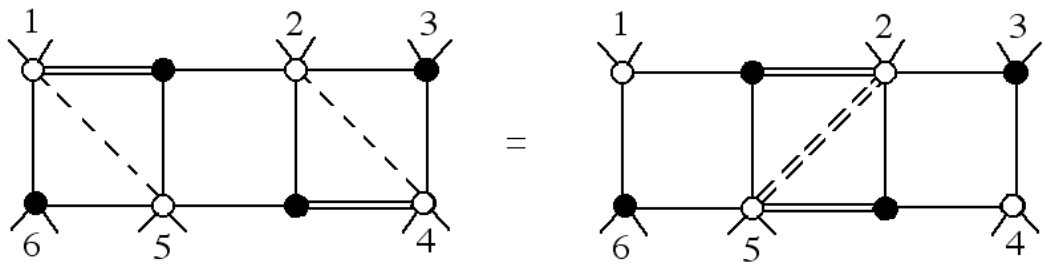

and the obvious dual of this. Another way of obtaining these six-field formulas is by the convolution of two five-field formulas into:

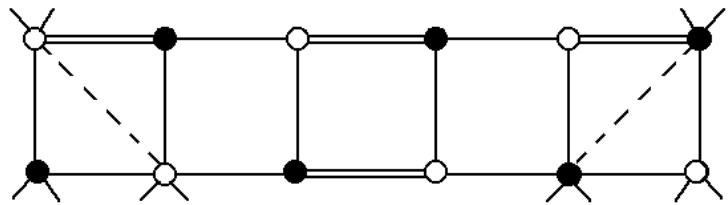

where the internal double lines effect an on-shell summation over scalar field states. The internal lines of the resulting diagram can then be telescoped by using

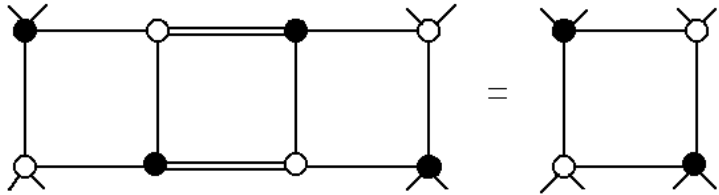

These six-field formulas can themselves used in convolution, and so lead to a more general δ-*function formula rule*. Suppose we have δ-function formulas $D_L$ and $D_R$ for $l$ scalar fields and for $r$ scalar fields respectively.

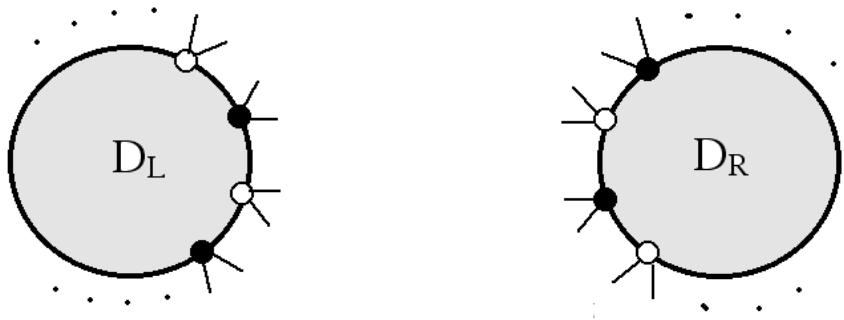

Then there are δ-function formulas for $l + r - 2$ scalar fields given by:

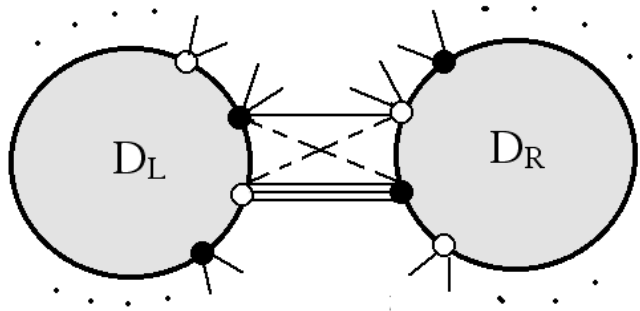

and also by

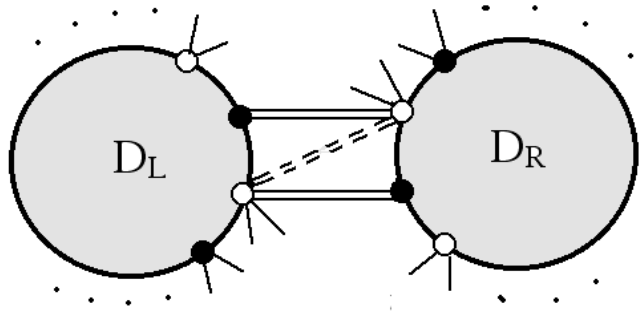

and its dual.

## 4. MHV Amplitudes

We can use the last of these δ-function formulas to write down all MHV amplitudes as twistor diagrams. First consider amplitudes of form $A(--+++...++)$, with Parke-Taylor formula

$$\frac{\langle 12 \rangle^3}{\langle 23 \rangle \langle 34 \rangle \langle 45 \rangle ... \langle n-1, n \rangle \langle n1 \rangle} \delta(p_1 + p_2 + p_3 + ... + p_n)$$

It may be seen, by an extended integration by parts generalising the four-field case, that agreement with the Parke-Taylor formula arises from a diagram of the following form:

For *n* even:

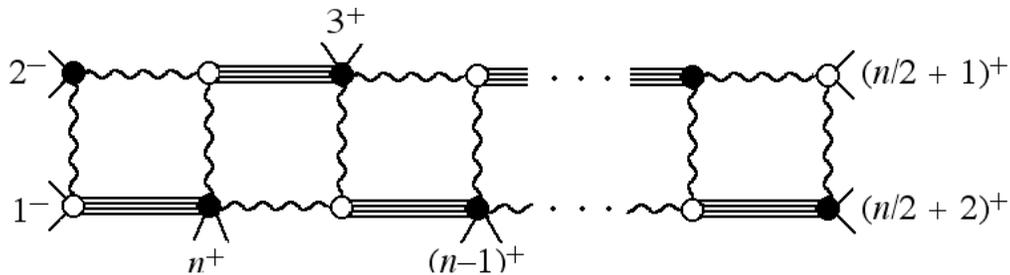

and for *n* odd:

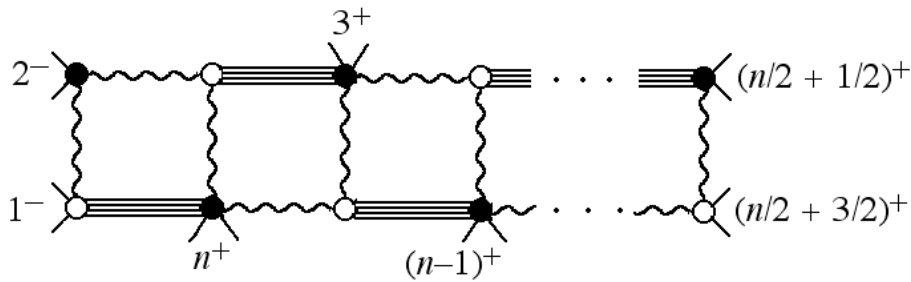

When the negative-helicity fields are not contiguous they must appear thus:

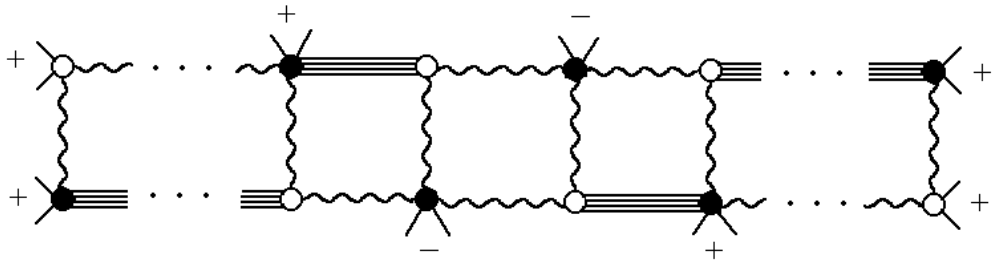

The anti-MHV case A(+ + − − − ... − − ) obviously has a dual representation.

Again, no purpose is served by trying to put momentum states into the diagram. Knowledge of the δ-function representation is all that is needed. Actually, the diagrams can if desired be seen as *encoding* the conventional momentum-space formulas in a combinatorial form which brings their conformally invariant aspects to light. To decode a diagram we need only relate it to its corresponding δ-function formula through the operators that have been described. This decoding relation is purely algebraic and combinatorial. One could use the formalism for this encoding purpose without knowing anything about the compact contour structure or considering amplitudes for any actual finite-normed fields.

It is a striking emergent feature of the results that the gauge-theoretic trace always runs along the loop formed by the exterior of the twistor diagram. Thus, it is natural to think of the diagram as giving a concrete realization of a 'ribbon' diagram, in which a gauge-theoretic index runs round the exterior lines, joining the exterior fields in the appropriate way.

Given an amplitude, there is in general more than one way of expressing it as a twistor diagram. As an example, the $A(1^- \; 2^- \; 3^+ \; 4^+ \; 5^+)$ amplitude can be represented in five different ways as

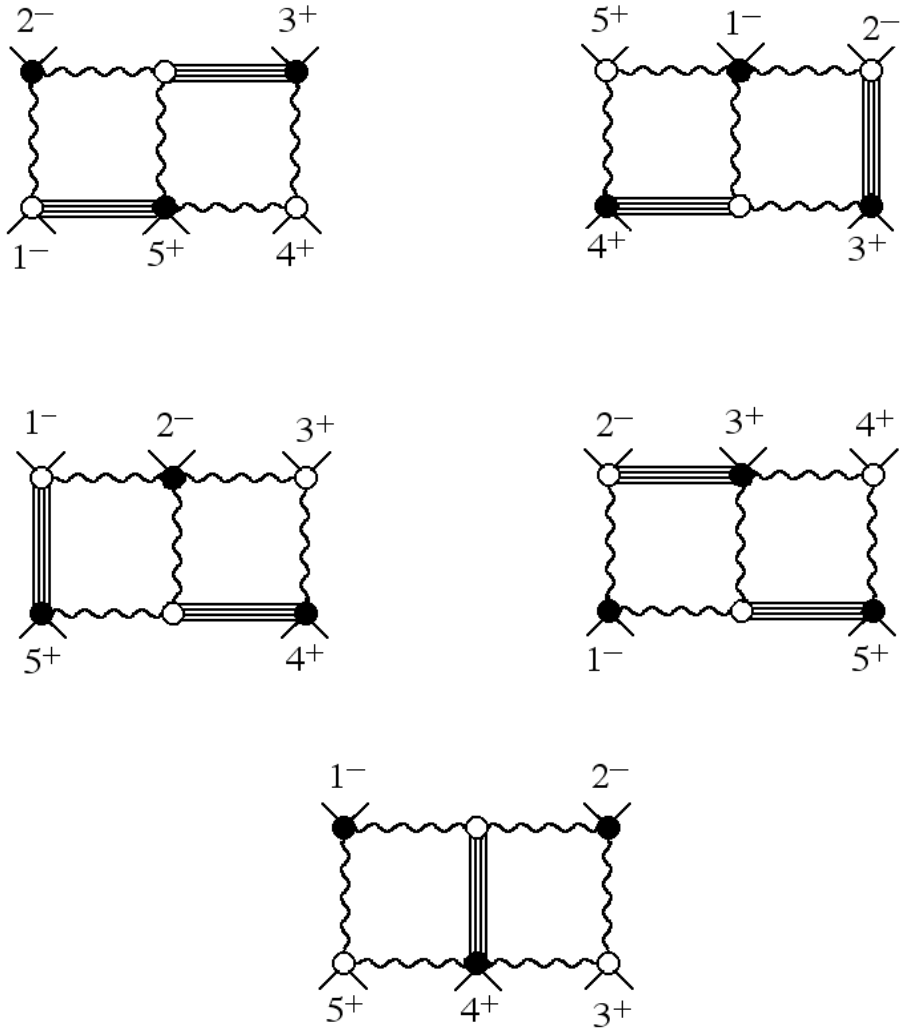

These diagrams should be interpreted as *representing* some deeper twistor-geometric entity. In particular, there are identities and linear dependences between diagrams which differ only in their boundary lines. This is only to be expected as such identities and linear relations simply reflect linear relations between corresponding contours: i.e. the pure geometry of twistor space.

We shall be able to characterise this multiplicity of representation in another way, by applying the result which will be obtained in the next section.

## 5. Diagram Bridging

The rule defined and proved by Britto Cachazo and Feng (2004) works for each colour-ordering independently, and it will be assumed henceforth that external gauge fields labelled 1, 2... *n* are in cyclic order. Their structure may be described as one of building a *bridge* between sub-processes (the bridge is, of course, an off-shell gauge field). We select two consecutive elements as *bridge-ends*. One must be of positive and one of negative helicity. This is no loss of generality, since there must be at least one such consecutive pair in a non-zero amplitude. Let us suppose, again without loss of generality, that the bridge-end fields are $x^+$ and $y^-$, where $y = x + 1$, modulo *n*. All additions and summations are implicitly modulo *n*.

Then the BCF recursive formula is:

$A(1, 2, 3... x - 1, x, y, y + 1... n - 1, n) =$

$$\sum_{i=y+1}^{x-2} A_L^+(i+1, i+2...x-1, \hat{x}^+, \hat{k}^+) \frac{1}{P^2} A_R^-(\hat{k}^-, \hat{y}^-, y+1...i-1, i) \ +$$

$$\sum_{i=y+1}^{x-2} A_L^-(i+1, i+2...x-1, \hat{x}^+, \hat{k}^-) \frac{1}{P^2} A_R^+(\hat{k}^+, \hat{y}^-, y+1...i-1, i)$$

where here the various A are actually the coefficients of the appropriate δ-functions of momenta, rather than the complete amplitudes. The 'hatted' variables and the $P^2$ will be described a little later. Our contention is that this bridging procedure is equivalent to a rule for bridging twistor diagrams. We assume, as an inductive hypothesis, that we have at our disposal twistor diagram representatives of all amplitudes of order up to (*n* − 1). In what follows we shall need four of these, corresponding to the sub-amplitudes $A_L^+$, $A_R^-$, $A_L^-$, $A_R^+$ above. (Of course, the non-uniqueness of diagrams means that there is in general a choice as to which diagram to use for a particular sub-amplitude.) Suppose that our diagrams are of the form:

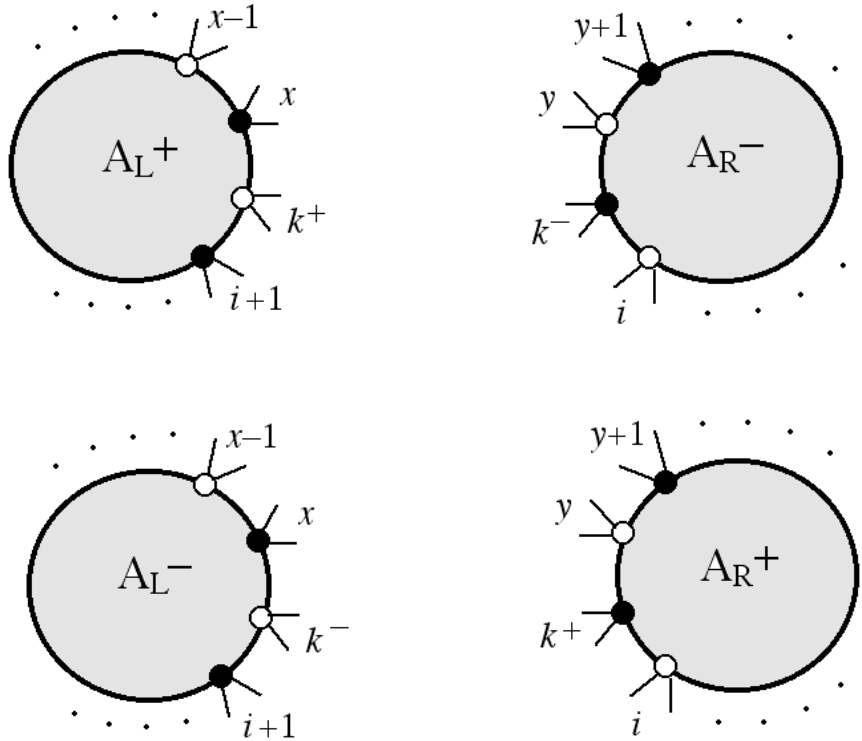

We are mainly concerned with the fields which surround the chosen 'bridge'. These are labelled $x$, $y$, $k^-$ and $k^+$. We shall refer to their corresponding twistor or dual twistor variables as $Z_x$, $W_y$, $Z_k$, $W_k$ in the obvious sense. Note that $A_L{}^+$ and $A_L{}^-$ are entirely different and independent diagrams. In $A_L{}^+$ the external function of $W_k$ is of homogeneity degree (–4); in $A_L{}^-$ it is of degree 0. Likewise $A_R{}^+$ and $A_R{}^-$ are different.

There is no significance at all to the fact that a dual twistor representation has been drawn for the fields $i$ and $(x - 1)$ and a twistor representation for fields $(y + 1)$ and $(i + 1)$. These are drawn in purely to help visualise the cyclic order, which is, of course, crucial.

We *have* made a choice about the $Z_x$, $W_y$, $Z_k$, $W_k$ variables which form the 'bridge', but this choice has no deep significance because twistor transforms can always be used to change the representation.

Our principal assertion can now be made. It is that the *i*th contribution to the complete *n*-field amplitude is given by the sum of the two twistor diagrams:

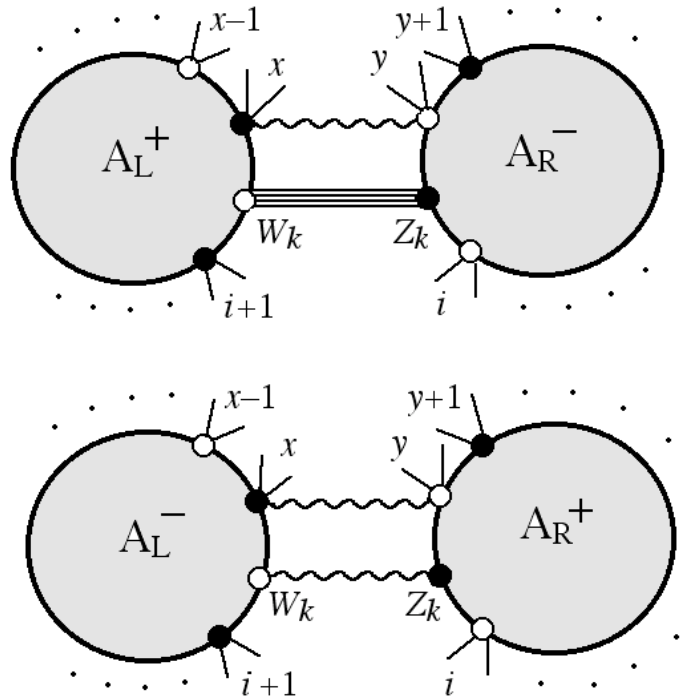

Note that the cyclic order (1, 2, 3... $x - 1$, $x$, $y$, $y + 1$... $n - 1$, $n$) is conserved in the resulting 'bridged' diagrams.

To show why this is, we work by analogy with the four-field and MHV examples. In those cases we proceeded by operating on the twistor diagram with differential operators corresponding to the *inverses* of the factors in the Parke-Taylor formulas. We then showed that a correct δ-function formula resulted. We shall do the same here. Our final step will, of course, use the δ-function rule arrived at in §3, which is based on having two δ-function diagrams and joining them together. So we need to show that the *inverses* of the factors in the the BCF formulas will convert the two sub-amplitude parts of our bridged diagram into correct δ-function formulas, and also produce the correct form of bridge between the two.

The key idea is that the 'hatted' spinors as defined by Britto Cachazo and Feng are exactly defined so as to correspond to this twistor diagram operation. To make the connection, recall that all the spinor objects used in momentum-space formulas appear in the twistor diagram as combinations of

$$I^{\alpha\beta} W_{r\beta}, \quad I_{\alpha\beta} Z_s^\beta, \quad I_{\alpha\beta} \partial/\partial W_{r\beta}, \quad I^{\alpha\beta} \partial/\partial Z_s^\beta$$

We shall use the shorter notation $W_r^A$, $Z_{sA'}$, $\partial W_{rA'}$, $\partial Z_s^A$ for these.

We continue by putting the essential ideas in terms of these differential operators. First note that the momentum corresponding to an external field described by twistor variable $Z_s$ or dual twistor variable $W_r$ is just

$$Z_{sA'} \partial Z_s^A \quad \text{or} \quad -W_r^A \partial W_{rA'}$$

so that the important sum

$$P^a = \sum_{j=i+1}^{x} p_j^a = -\sum_{j=y}^{i} p_j^a$$

which is the momentum which 'crosses the bridge', can readily be written down as a twistor operator. Applying integration by parts, or equivalently, using momentum conservation, this operator can be rewritten as

$$\partial Z_x^A Z_{xA'} - W_k^A \partial W_{kA'} \quad \text{or} \quad \partial Z_k^A Z_{kA'} - W_y^A \partial W_{yA'}$$

acting on the two lines of the bridge. The d'Alembertian operator $P^2$ is likewise

$$-2 W_k^A \partial Z_{xA} Z_{xA'} \partial W_k^{A'} = -2 W_y^A \partial Z_{kA} Z_{kA'} \partial W_y^{A'}$$

Now consider the effect of $\partial Z_x^A$. If it were left unmodified in the bridged diagrams, it would act on the boundary line joining it to $W_y$, and this would prevent the sub-amplitude twistor diagrams being correctly reduced to their corresponding δ-function formulas. So it must be modified in such a way that it becomes 'blind' to the bridge. The modification is:

$$\partial \hat{Z}_x^A = \partial Z_x^A - \tfrac{1}{2} \frac{P^2 W_y^A}{W_y^B P_{BA'} Z_x^{A'}} = \partial Z_x^A - W_y^A \frac{Z_k^{A'} \partial W_{yA'}}{Z_k^{B'} Z_{xB'}}$$

and similarly

$$\partial \hat{W}_{yA'} = \partial W_{yA'} - \tfrac{1}{2} \frac{P^2 Z_{xA'}}{W_y^A P_{AB'} Z_x^{B'}} = \partial W_{yA'} - Z_{xA'} \frac{W_k^A \partial Z_{xA}}{W_k^B W_{yB}}$$

These operators, acting on the two lines of the bridge, vanish. They will therefore behave within the bridged diagrams exactly as $\partial Z_x^A$, $\partial W_{yA'}$ did in the separate sub-amplitudes. Note that the definition using $P$ determines these operators in terms of the actual external fields of the bridged diagram.

Likewise, the $\hat{k}$ operators must be such as to act on the sub-amplitude parts of the bridged diagrams just as the $k$ operators did within the separated components. Moreover, $\partial \hat{W}_{kA'}$ must be defined so that it is equivalent to $\hat{Z}_{kA'}$ and $\partial \hat{Z}_{kA}$ likewise equivalent to $\hat{W}_{kA}$. Before defining them, note that:

$$\partial Z_k^A = W_k^A \frac{Z_{xA'} \partial W_k^{A'}}{Z_{xB'} Z_k^{B'}} \qquad \text{and} \qquad \partial W_{kA'} = Z_{xA'} \frac{W_y^A \partial Z_{kA}}{W_y^B W_{kB}}$$

We also have: $\qquad W_{kA} \partial W_{kA'} = Z_{xA'} \partial Z_{kA}$

Following Britto Cachazo and Feng, we can satisfy all these demands by:

$$\partial \hat{W}_{kA'} = \hat{Z}_{kA'} = \frac{P_{AA'} W_y^A}{W_y^B P_{BB'} Z_x^{B'}} = \frac{Z_{kA'}}{Z_{kB'} Z_x^{B'}}$$

$$\partial \hat{Z}_{kA} = \hat{W}_{kA} = P_{AA'} Z_x^{A'} = W_{kA} Z_{xA'} \partial W_k^{A'}$$

This looks asymmetric but it works because $\hat{Z}_k$ and $\hat{W}_k$ appear in the sub-amplitude formulas *either* in the combinations $\hat{W}_{kA} \partial \hat{W}_{kA'}$ or $\hat{Z}_{xA'} \partial \hat{Z}_{kA}$ (i.e. as momenta) *or* in terms which are homogeneous in $\hat{Z}_k$ and $\hat{W}_k$. In the first case we find, using the relations already noted,

$$\hat{W}_{kA} \partial \hat{W}_{kA'} = W_{kA} Z_{kA'} \frac{Z_{xB'} \partial W_k^{B'}}{Z_{xC'} Z_k^{C'}} = Z_{xA'} \partial Z_{kA} = W_{kA} \partial W_{kA'} = \hat{Z}_{kA'} \partial \hat{Z}_{kA}$$

so that the hatted terms do indeed act within the bridged diagram just as if there were external fields attached to $W_k$, $Z_k$. In the second case, the rescaling plays a role, but all that matters is the overall homogeneity of $\hat{W}_k$ and $\hat{Z}_k$: the asymmetry in the definition is not relevant. In both terms of the sum, the effect of the resulting factor is to change the original bridge to a 'scalar bridge':

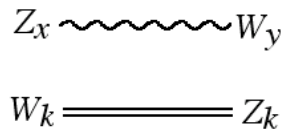

Finally, the operator $P^2$ is just the operator which transforms this scalar bridge to

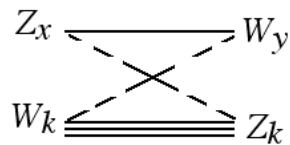

Now that the twistor equivalents of the BCF hatted spinor definitions are known, and their properties noted, we can apply the strategy as described above to show the correspondence between the BCF formula and the bridged twistor diagrams. Thinking of the formula as defining operators applied to the δ-function, we apply the corresponding *inverse* operators to the bridged twistor diagrams. These operators are such that they act on the two components just as if the bridge were not there, except that the bridge is transformed into just the correct form for the application of the δ-function rule. We deduce that the whole expression, operating on the bridged diagram, will produce a δ-function formula for *n* fields. But this is exactly what we require to establish the result claimed.

One important element remains. For the application of the BCF formula we need expressions for *three*-field-amplitudes, even though such amplitudes are not properly defined for actual quantum states. The twistor diagram equivalents are simply:

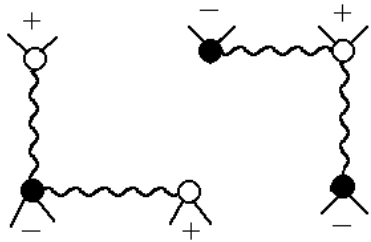

However, it is essential to add an *extra rule:* the three-field amplitude *vanishes* if the variable being integrated out is of the same helicity type as the one not being used as a bridge-end.

## 6. Examples of diagram bridging to find non-MHV amplitudes

It is now straightforward to write down the terms arising in the application of the BCF formula. It is useful to verify the production of 4 and 5-field MHV processes from 3-vertex terms. This is instructive because we learn from this that the ambiguity of expression in these amplitudes is exactly accounted for by the different possible choices of where to build the bridges. However, we will go straight to the non-MHV processes.

For $A(1^- \ 2^- \ 3^- \ 4^+ \ 5^+ \ 6^+)$ we choose the 'bridge' to be between $x = 6$ and $y = 1$ (thus following Britto Cachazo and Feng (2004), for the sake of easy comparison with their formulas).

Of the six terms which arise, four vanish immediately because $A(1^- \ 2^- \ k^-)$, $A(1^- \ 2^- \ 3^- \ k^-)$, $A(1^- \ 2^- \ 3^- \ k^+)$ and $A(k^+ \ 5^+ \ 6^+)$ vanish. Non-zero terms arise from the composition of:

(a) $A(1^- \ 2^- \ k^+) \circ A(k^- \ 3^- \ 4^+ \ 5^+ \ 6^+)$

(b) $A(1^- \ 2^- \ 3^- \ 4^+ \ k^+) \circ A(k^- \ 5^+ \ 6^+)$.

For (a) we can take as diagrams for the two sub-amplitudes:

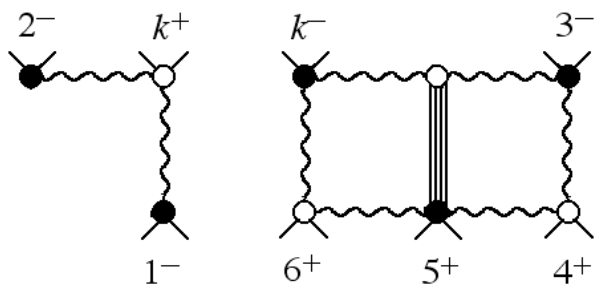

We have made a particular choice (of no significance) of the five possible diagrams for the 5-field MHV sub-amplitude. Now, we apply the diagram bridging rule and, hey presto, we obtain the diagram:

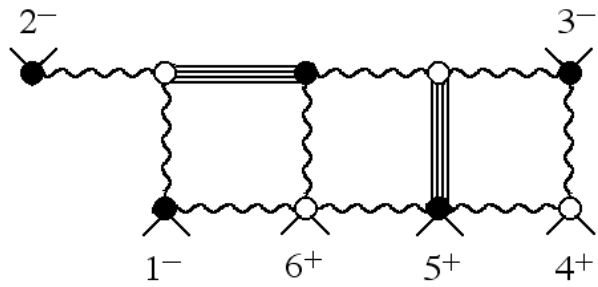

The second term is just the dual of this, so we arrive at the complete result

A(1− 2− 3− 4+ 5+ 6+ ) =

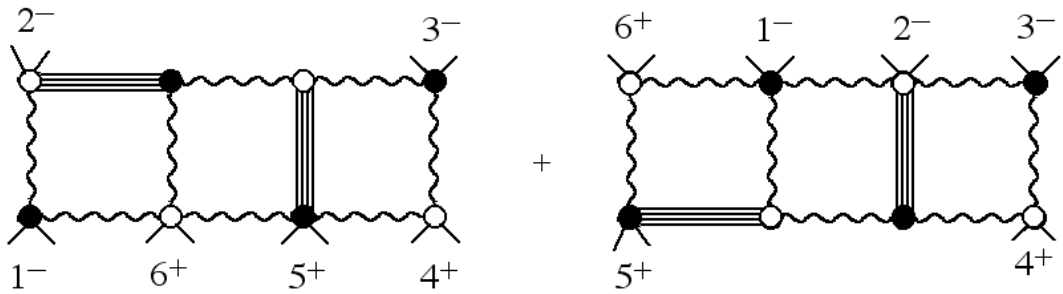

In alternative representations for the complete amplitude, the first of these terms could be represented as (for instance) either of these diagrams:

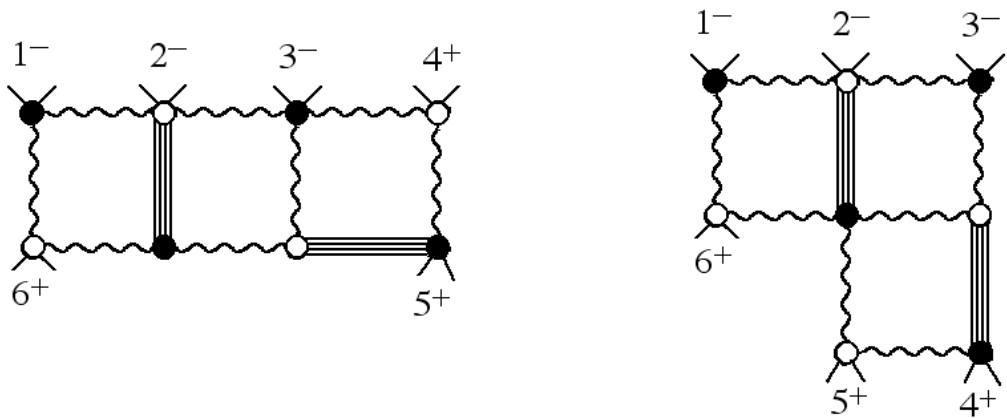

Note how in every case the form being integrated is exactly the same: only the location of the boundaries is changed from one diagram representation to another.

For A(1− 2+ 3+ 4− 5− 6+ ), again we bridge on (1− 6+).
The extra rule regarding 3-vertices means that there is no contribution from joining A(1− 2+ k+ ) A(k− 3+ 4− 5− 6+ ) or joining A(1− 2+ 3+ 4− k+ ) A(k− 5− 6+ ).
We are left with three nonvanishing terms, and obtain the sum of:

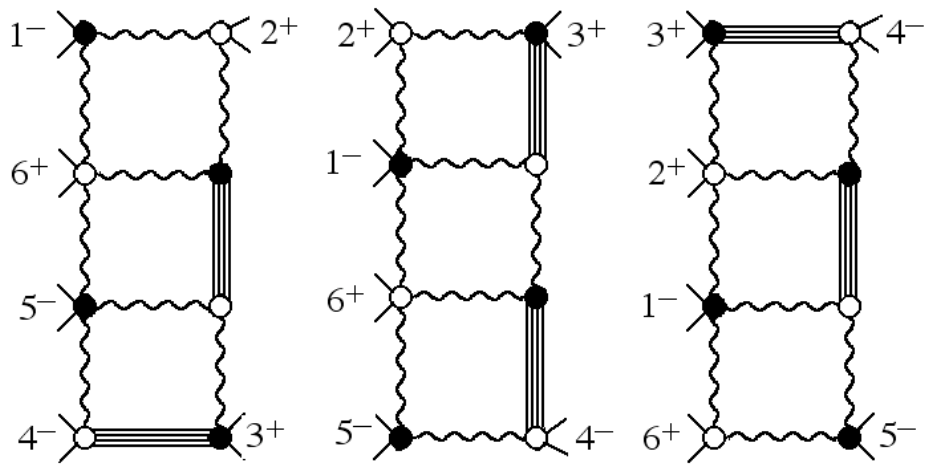

For A(1− 2+ 3− 4+ 5− 6+ ) similarly we obtain the sum of:

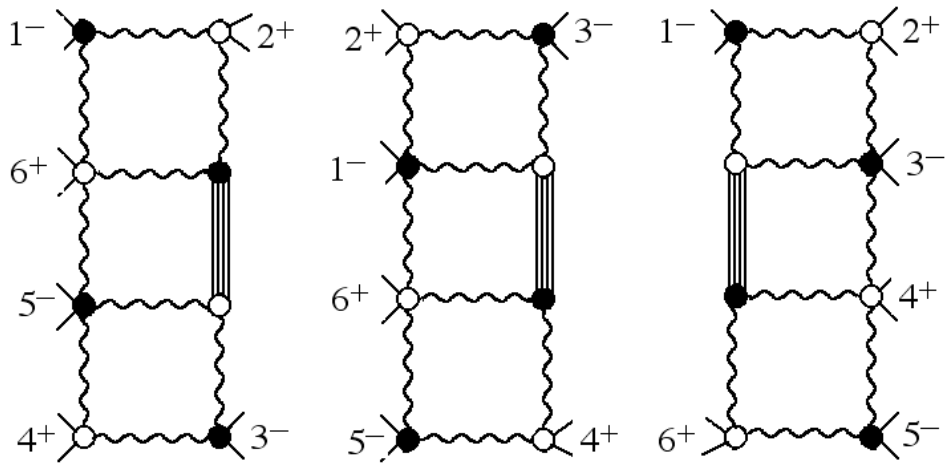

The usefulness of this representation has been verified by studying twistor diagrams for the seven-field and eight-field tree amplitudes computed by Britto Cachazo and Feng (2004), also by Roiban, Spradlin and Volovich (2004), Bern, Del Duca, Dixon, and Kosower (2004). In some cases, the diagram representation makes it easier to see the identity of various terms that arise. However, the important and difficult linear dependences between various formulas, as noted by these authors, cannot be be read off directly from this new diagram representation. We can only say that the geometry of the diagrams opens the door to a completely new description of such relationships, based on the pure homology theory (more accurately, the relative homology) of the integration space. We add a few remarks on this development in the next section.

## 7. Twistor Quilts

The striking geometric relationship of the diagram to the gauge-theoretic trace obviously suggests a relationship with *open strings*. (This connection was noticed long ago (Hodges 1990, 1998) but in woeful ignorance of the astonishing generalisation already effected by Parke, Taylor (1986) and others, its potential was not properly appreciated!) We are naturally led to the suggestion that the non-unique representation of amplitudes by diagrams can be understood in terms of these different but equivalent diagrams being merely different ways of dividing up an underlying string-like object. These divisions are not so much like *ribbons* as like *quilts*. It seems very likely that different 'quilts' for a given amplitude can be expressed entirely in terms of different choices of bridge-ends in applications of the bridging process.

A striking fact is that it is that it is not actually necessary to specify which lines represent the quadruple poles, and which are the boundaries. If an external field is of homogeneity (–4), it 'forces' the lines meeting its vertex to be boundary lines, and these in turn force the others. It will be found that all the diagrams have sufficient external (–4) functions to determine the identity of *all* the internal lines. This feature of the diagrams seems to be intimately related to the rule about vanishing 3-vertices. It would also seem to be related to the very nature of tree diagrams. Just as Feynman diagrams for tree amplitudes carry *momenta* which are fully determined by the external fields, so tree twistor diagrams carry *helicities* which are likewise uniquely specified. (*Ipso facto*, we expect this property to *fail* for loop diagrams.)

In these tree diagrams one could actually *replace* each quadruple pole by a boundary line, together with the numerical factor $(24\ k^{-4})$, and the result of the integration would be the same. In this way, *everything* in the diagram becomes pure geometry. This again strongly suggests that identities and linear relationships between amplitudes are equivalent to purely geometrical relationships between homology classes.

To specify a twistor diagram, then, it suffices to list the ordering of vertices and the input of external fields. We can illustrate this by drawing streamlined 'quilt' diagrams which emphasise the purely geometrical characteristics. Each line in a 'quilt' simply represents a boundary subspace of form $W_\alpha Z^\alpha = k$. The five different but equivalent diagrams for A(1− 2− 3+ 4+ 5+), for instance, can be written:

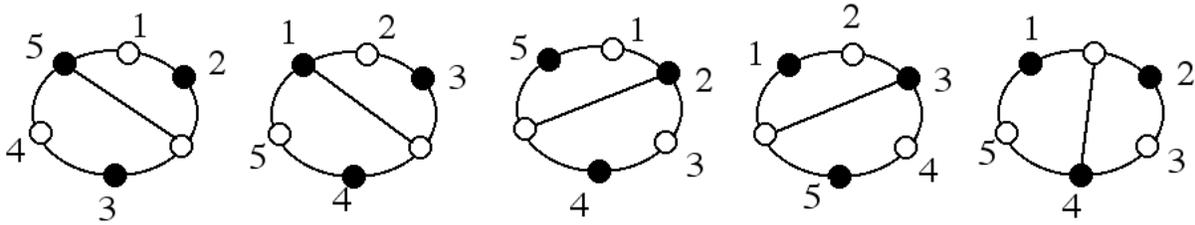

As a more complicated illustration, we can express the linear relationship needed by Britto Cachazo and Feng to demonstrate the symmetry of their sum for A(1+ 2− 3+ 4− 5+ 6− 7+ 8− ) thus:

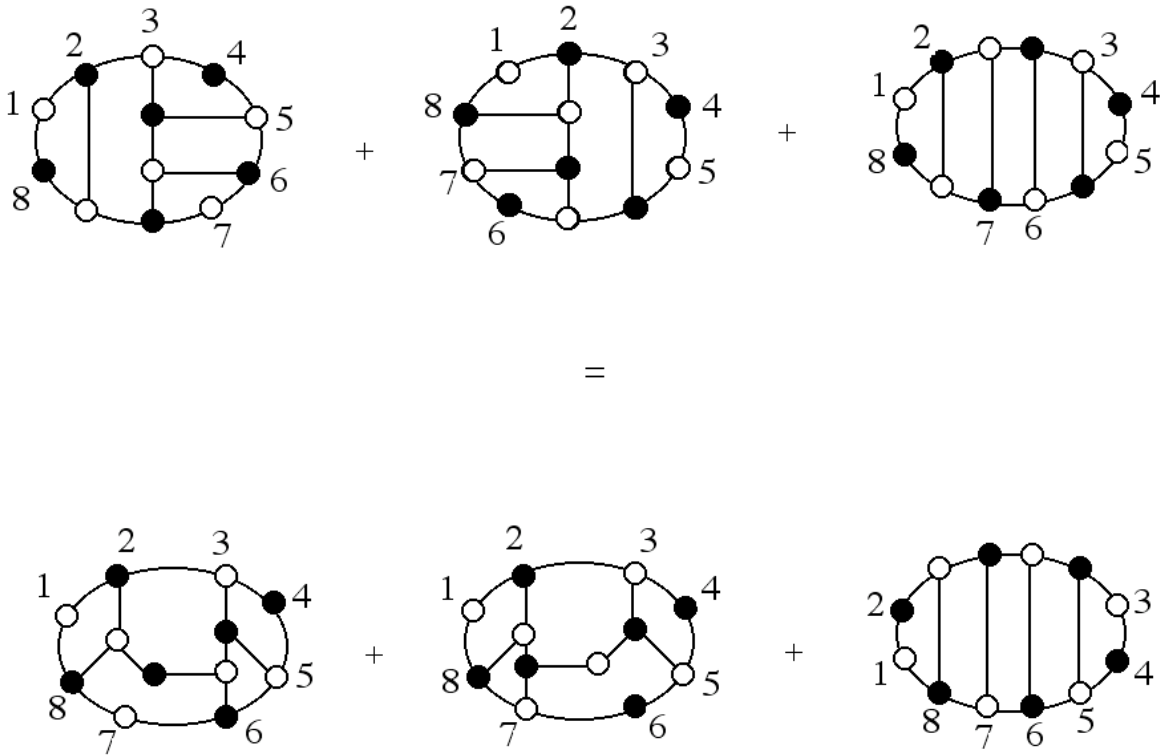

These six 'quilts' correspond to the expressions described by Britto Cachazo and Feng as $W, g^5 rW, X, g^2 T, g^7 T, g^7 V$ respectively. Indeed they encode all the information in those expressions. It seems possible that this will lead to useful combinatorial methods for establishing identities.

Obviously, the diagrams are helpful in seeing discrete symmetries, but it is perhaps even more important that they bring to light the essential elements of *conformal* symmetry which have not hitherto been emphasised. The apparently weird rational functions of momenta which appear in the long formulas for these amplitudes, can surely now be better understood as the outcome of applying conformally invariant operators — a tightly defined range of combinatorial and algebraic ingredients.

But perhaps the most vivid feature of the twistor diagrams is that they tell a *physically significant story* about the process. The story lies in the 'cut' structure, where we immediately see the collinear and multiparticle singularities corresponding to physically possible sub-processes. It should be possible also to derive a simple rule for what happens when an external gauge-field is 'dropped', since rough inspection shows that if external fields of homogeneity degree 0 are simply chopped off, a new contour emerges which yields the correct amplitude for the corresponding process with one less field. These reduction phenomena are in themselves strong constraints on the forms that the diagrams can take, and suggest that useful theorems can be gained by combining them with analytic S-matrix insights.

## 8. Extension to loop diagrams

Can these results be extended to loop diagrams? There are encouraging signs. Although the focus of this note has been on gauge fields, twistor diagrams are equally applicable to other zero-rest-mass field theories. It is straightforward to write down tree diagrams for massless scalar field theory, and there are obvious candidates (as yet unverified) for structures corresponding to one-loop scalar diagrams. If this can be verified, it should be possible to use a twistor diagram version of the 'box-function' as a template for the gauge-theoretic one-loop diagrams in analogy with the use of the simple δ-function scalar diagrams for tree diagrams, and so obtain analogous one-loop gauge-theoretic expressions. It is noteworthy that the amplitude formulas obtained by 'cutting' loop diagrams do fit with the twistor-diagram representation, and this again suggests there is something important to be found at the loop level. It is also noteworthy that the twistor diagrams have so strong a connection with 'cut' structure.

However, considerable caution is required. We do not know which theory we expect to find at this level: N=0 or N=4 or something else again? It is not clear whether the ultra-violet divergence structure can be handled correctly within the formalism, although for infra-red divergences, the structure already discussed at tree level seems very promising. And there will in general be information in Feynman loop diagrams that does *not* appear in their cut structure: we have no evidence that any such information will be correctly supplied by a twistor-diagram formalism.

An interesting possibility is that for loop-like diagram lines, where helicities are not determined by external fields, the inhomogeneity will liberate the twistor line-propagators from interpretation in terms of helicity eigenstates and give something quite new.

## 9. Further implications

If we study the the *internal vertices* in these twistor diagrams, we note an internal twistor vertex for every departure from MHV. Dually, there is a dual-twistor internal vertex for every departure from anti-MHV. There seems to be a definite connection with the pictures called 'twistor diagrams' in the theory of Witten (2003), showing how external fields can be considered as confined to lines, or more generally several intersecting lines, in twistor space, with the number of lines determined by the departure from MHV.

Edward Witten's theory has already led to much greater understanding of the twistor diagram structure for tree amplitudes, and how this should be related to one-loop diagrams. There is doubtless much more to be learnt.

The correspondence with Witten's theory is, however, indirect. If our line-propagators were all simple poles then they could be thought of as restricting all the variables involved to certain lines in twistor space. But instead, they are the third derivatives of, or the inverse derivatives of, such poles. Another, more radical, difference lies in the inhomogeneity employed in our formalism. Thus the structures remarked on in this note cannot simply be deductions from Witten's theory.

Finally, the diagram recursion principle gives a clear indication of an *autonomous generating principle*. The recursion relation can be put in terms of rules for the composition of 3-field amplitudes. It should be likewise possible to derive rules for generating twistor diagrams from formal 3-vertices, in a Lagrangian-like form. The 'ribbon' aspect of the diagrams also suggests a connection with the beautiful geometric ideas of string theory which have inspired so many advances. In conclusion, I express indebtedness to the tireless ingenuity of the quantum field theorists who have revealed such astonishing structure in this formidable problem.

## Acknowledgements

I am grateful for the opportunity to discuss some of these ideas with Roger Penrose, Philip Candelas, Valentin Khoze, Lionel Mason, and David Skinner. I am also in debt to all those who organised and spoke at the conference on twistors and strings at Oxford University in January 2005, and so stimulated these new observations.

Further supporting material will appear on http://www.twistordiagrams.org.uk